\documentclass[sigconf,authorversion,nonacm]{acmart}
\AtBeginDocument{%
  \providecommand\BibTeX{{%
    \normalfont B\kern-0.5em{\scshape i\kern-0.25em b}\kern-0.8em\TeX}}}

\setcopyright{acmlicensed}
\copyrightyear{2018}
\acmYear{2018}
\acmDOI{XXXXXXX.XXXXXXX}

\acmConference[Conference acronym 'XX]{Make sure to enter the correct
  conference title from your rights confirmation emai}{June 03--05,
  2018}{Woodstock, NY}
\acmISBN{978-1-4503-XXXX-X/18/06}

\begin{document}

\title{CoSearchAgent: A Lightweight Collaborative Search
Agent with Large Language Models}

\author{Peiyuan Gong}
\affiliation{%
  \institution{GSAI, Renmin University of China}
  \city{Beijing}
  \country{China}}
\email{pygongnlp@gmail.com}

\author{Jiamian Li}
\affiliation{%
  \institution{GSAI, Renmin University of China}
  \city{Beijing}
  \country{China}}
\email{miankora33@ruc.edu.cn}

\author{Jiaxin Mao}
\affiliation{%
  \institution{GSAI, Renmin University of China}
  \city{Beijing}
  \country{China}}
\email{maojiaxin@gmail.com
}

\renewcommand{\shortauthors}{Trovato and Tobin, et al.}

\begin{abstract}
Collaborative search supports multiple users working together to accomplish a specific search task.
Research has found that designing lightweight collaborative search plugins within instant messaging platforms aligns better with users' collaborative habits.
However, due to the complexity of multi-user interaction scenarios, it is challenging to implement a fully functioning lightweight collaborative search system. 
Therefore, previous studies on lightweight collaborative search had to rely on the Wizard of Oz paradigm.
In recent years, large language models (LLMs) have been demonstrated to interact naturally with users and achieve complex information-seeking tasks through LLM-based agents.
Hence, to better support the research in collaborative search, in this demo, we propose \textbf{CoSearchAgent}, a lightweight collaborative search agent powered by LLMs.
CoSearchAgent is designed as a Slack plugin that can support collaborative search during multi-party conversations on this platform. Equipped with the capacity to understand the queries and context in multi-user conversations and the ability to search the Web for relevant information via APIs, CoSearchAgent can respond to user queries with answers grounded on the relevant search results. It can also ask clarifying questions when the information needs are unclear. The proposed CoSearchAgent is highly flexible and would be useful for supporting further research on collaborative search.
The code\footnote{https://github.com/pygongnlp/CoSearchAgent} and demo video\footnote{https://github.com/pygongnlp/CoSearchAgent/blob/master/demo.mp4} are accessible.

\end{abstract}







\maketitle

\begin{figure}[htbp]
    \centering
    \includegraphics[width=\linewidth]{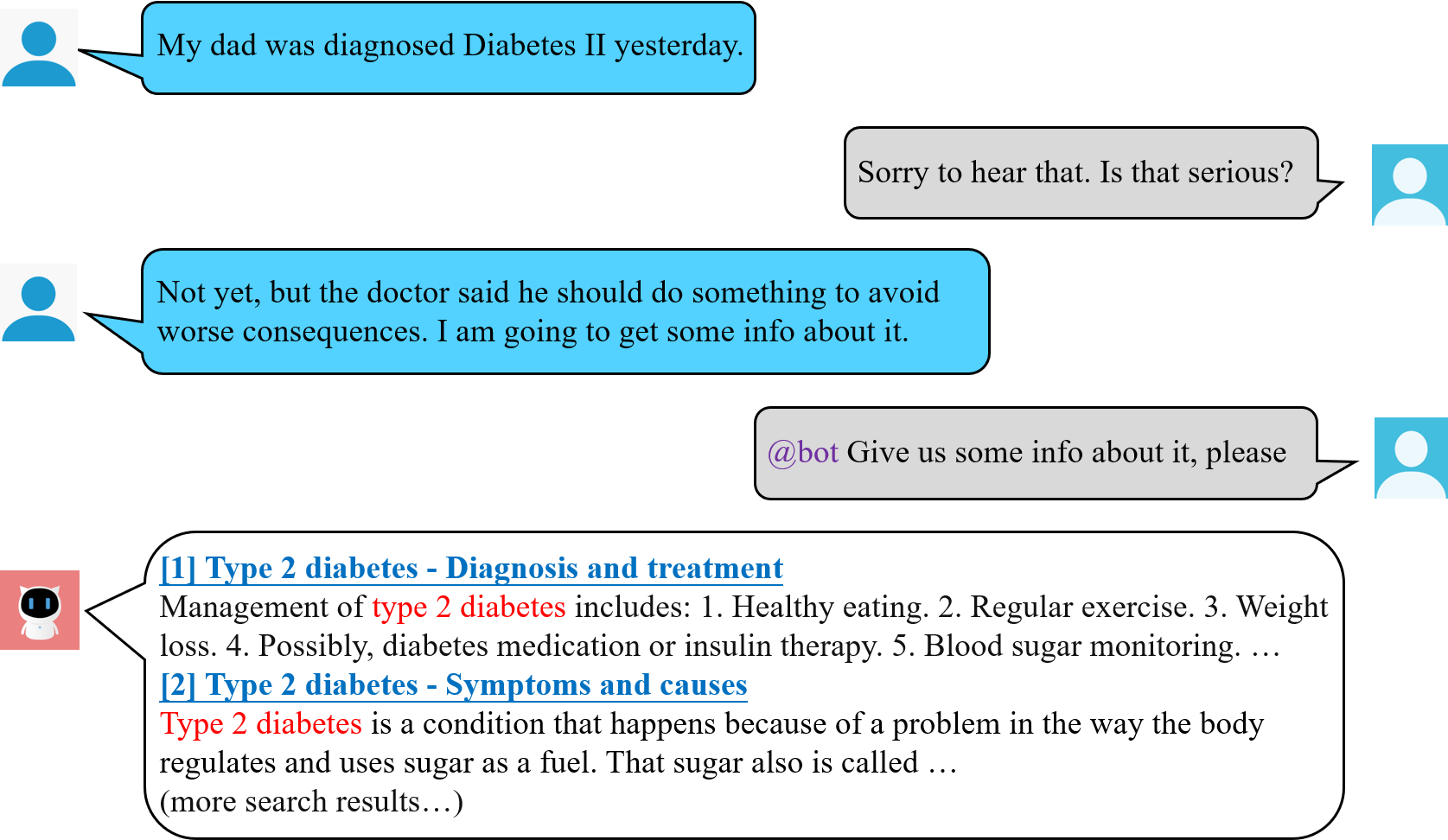}
    \caption{An instance of lightweight collaborative search. The collaborative search system needs to understand the conversational context of interactions between two users and propose search results to the users.}
    \label{fig:intro}
\end{figure}

\vspace{-3mm}
\section{INTRODUCTION}
Collaborative search has become a prominent topic in the field of information retrieval in recent years
\cite{shah2010collaborative, shah2014collaborative}. It involves addressing shared search needs among multiple users, fostering discussions on complex search goals, enhancing understanding of others' search behaviors, and facilitating the exchange of search results.
\citet{morris2013collaborative} shows that many people are accustomed to engaging in collaborative searches at least once a week, and this frequency is increasing year by year.
Hence, designing user-friendly and powerful collaborative search systems has become increasingly important.

The most common research paradigm in collaborative search is to build dedicated search software \cite{paul2009cosense, amershi2008cosearch, morris2007searchtogether, putra2018searchx}.
This type of software offers core functionalities, including search and chat.
Users can engage in conversations within the chat box and submit queries in the search box.
However, \citet{morris2013collaborative} demonstrates that in comparison to using dedicated collaborative search software, users are more accustomed to separately using instant messaging platforms and search tools to accomplish tasks.

Therefore, currently, research efforts have focused on integrating collaborative search systems into communications on instant messaging platforms, a concept known as lightweight collaborative search \cite{avula2018searchbots, avula2019embedding}.
As illustrated in Figure \ref{fig:intro}, during a discussion between two users, the collaborative search plug-in seamlessly integrates into the conversation and proposes relevant search results to them.
Moreover, several works have explored mixed interaction approaches in this scenario \cite{avula2022effects, avula2023and}, such as asking clarifying questions or providing search suggestions.
However, due to the intricacies of multi-user interaction scenarios, implementing a fully functional lightweight collaborative search system is challenging.
Therefore, earlier studies frequently employ a Wizard of Oz approach to simulate genuine collaborative search systems \cite{avula2018searchbots, avula2019embedding, avula2020wizard, avula2022effects, avula2023and}.

In this demo, we present CoSearchAgent, a lightweight collaborative search agent powered by large language models (LLMs), harnessing the robust abilities of LLMs in understanding instructions and engaging interactively \cite{wei2022emergent, wang2023survey, zhao2023survey}. 
To the best of our knowledge, we are the first to utilize LLMs in the collaborative search scenario.
CoSearchAgent is crafted as a Slack plugin to facilitate collaborative search within multi-party conversations on the Slack platform. 
With the capability to comprehend queries and contexts in multi-user dialogues, as well as the ability to retrieve pertinent information from the Web through APIs, CoSearchAgent can furnish not just search results but also generate answers derived from those results to present to users.
It also supports mixed-initiative dialogue with users.
When seeing ambiguous queries, it asks clarifying questions to further specify user needs.
In addition, as an open-source collaborative search plugin, CoSearchAgent is highly customizable and supports recording users' conversations and interactions, such as search and click.
This capability empowers researchers to analyze user behavior across various collaborative search domains, thus fostering advancements in collaborative search research.

\section{RELATED WORK}
In recent years, with users' search goals becoming increasingly complex, research on developing collaborative search systems that can assist multiple users in searching collectively has become a hot topic \cite{shah2010collaborative, shah2014collaborative, morris2022collaborative}.
Currently, collaborative search can be categorized into two paradigms.
The first involves developing dedicated collaborative search software with an interface comprising two modules: search and chat. Users can communicate in the chat box, enter queries into the search box to obtain a list of search results, and share these results with other users \cite{amershi2008cosearch, morris2007searchtogether, putra2018searchx,shah2010collaborative}.
However, \citet{morris2013collaborative} shows that although the number of users engaging in collaborative searches increases annually, the majority are not accustomed to using proprietary systems. Instead, they opt to separately utilize instant messaging platforms and search engines to accomplish collaborative search tasks.

Another collaborative search paradigm addresses this issue by designing collaborative search plugins on instant messaging platforms such as Slack, thereby integrating them into users' conversations.
\citet{avula2018searchbots, avula2019embedding} embed search results into the multi-party conversational context, making them accessible for all users to review.
\citet{avula2022effects} investigates mixed interactive behaviors in collaborative search scenarios, including asking clarifying questions and providing search suggestions.
\citet{avula2023and} explores when and why a system should engage in proactive interactions.
Notably, due to the intricate nature of multi-user interactions, creating a fully functional lightweight collaborative search system is challenging. Therefore, previous studies have often applied the Wizard of Oz study \cite{avula2020wizard}.
In this demo, We implemented a practical collaborative search agent using LLMs, capable of participating in user conversations and offering necessary assistance.

\section{METHODOLOGY}
To illustrate why CoSearchAgent excels as an exceptional lightweight collaborative search system, in this section, we provide a comprehensive explanation of its multi-user interaction capabilities. 
These encompass processing queries through understanding the context of multi-person conversations, providing search results to users, and generating accurate answers with citation markers based on these search results.

\begin{figure*}
    \centering
    \includegraphics[width=\linewidth]{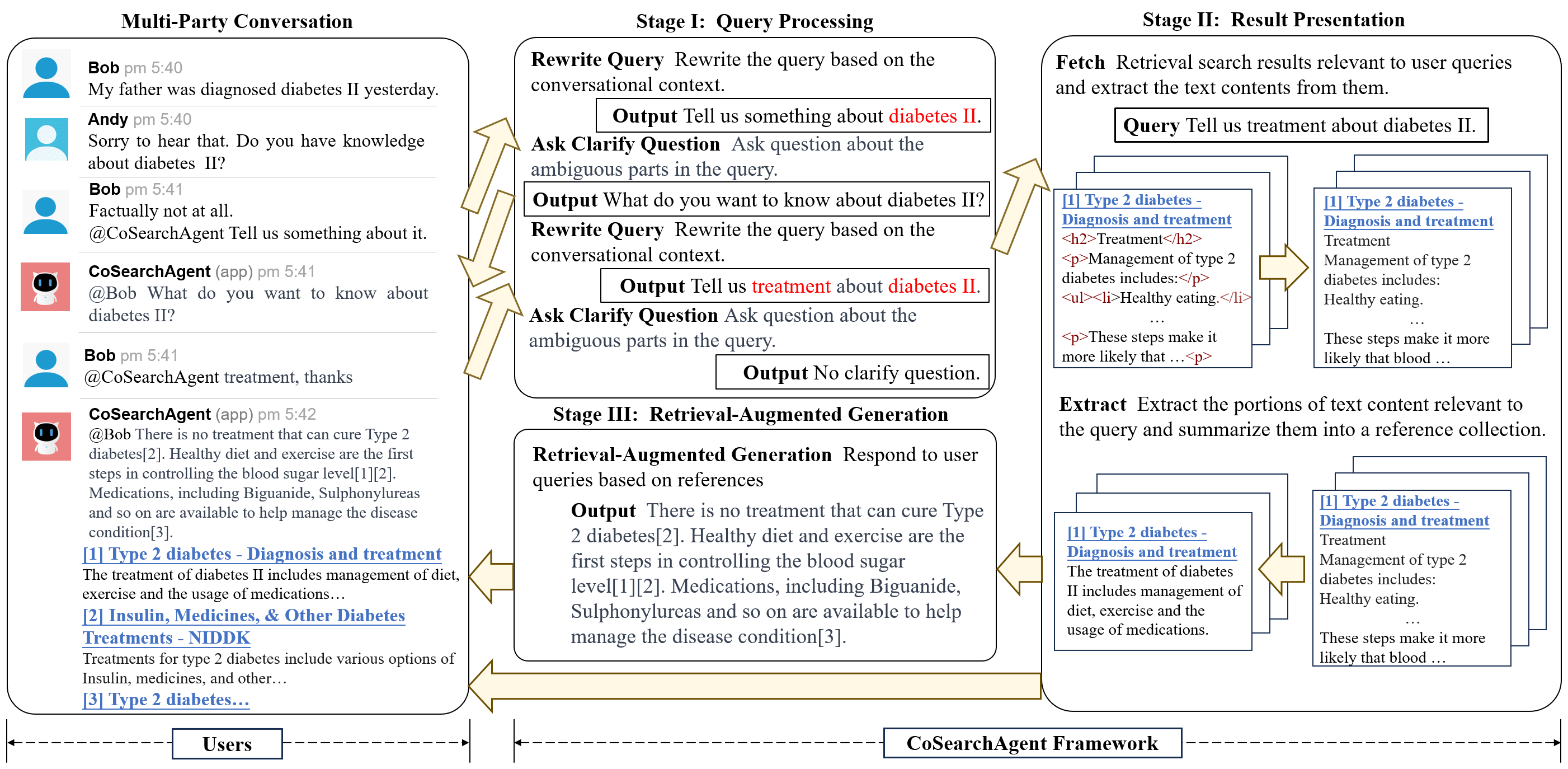}
    \caption{The overall architecture of CoSearchAgent. Given a context of multi-party dialogue and a query posed by a user, CoSearchAgent provides its response through the following three steps: \textbf{(I) Query Processing}: Rewriting the query based on the dialogue context and asking clarification questions for ambiguous parts of the query; \textbf{(II) Search Results Presentation}: Retrieve search results, extract relevant contents related to the query, and provide them to users as references; \textbf{(III) Retrieval-Augmented Generation}: Responding to the user's query relying on the generated references.}
    \label{collabsearchagent}
\end{figure*}

% \vspace{-2mm}
\subsection{Query Processing}
Due to the complexity of multi-party conversation scenarios and the relevance of user queries to the conversational context, previous studies on lightweight collaborative search have typically relied on the Wizard of Oz paradigm \cite{avula2018searchbots, avula2019embedding, avula2020wizard, avula2022effects, avula2023and}. This entails human operators reading the context of the conversation, inputting queries into the wizard of oz software for search\cite{avula2020wizard}, and returning search results to the chat box.
The emergence of LLMs may potentially replace this paradigm, as they possess strong interactive and comprehension capabilities \cite{zhao2023survey, wang2023survey, zhu2023large}.
As shown in Figure \ref{collabsearchagent},
to enable the CoSearchAgent to comprehend the context of multi-party conversations and address the user query, we utilize LLM to read the multi-party conversational context $U$, which is composed by $n$ utterances from different users, represented as $U = \{u_1, u_2, \ldots, u_n\}$, and then rewrite incomplete sections in the query and ask clarifying questions for the ambiguous portions after the rewrite.
% \vspace{-1mm}
\subsubsection{Rewrite query} 
CoSearchAgent can rewrite incomplete portions of user queries based on conversational contexts \cite{yu2020few, ye2023enhancing, mo2023convgqr}.
With the task instruction $F_{rewrite}$ for query rewriting, considering the multi-party conversational context $U$ and a query $q$, CoSearchAgent is expected to rephrase the incomplete segments within the query for search effectively:
\begin{equation}
    q_{rewrite} = F_{rewrite}(U, q)
\end{equation}
Where $q_{rewrite}$ denotes the revised form of the query $q$.
If there are no incomplete parts in the query, then $q_{rewrite} = q$.
% \vspace{-2mm}
\subsubsection{Ask clarifying question}
User queries may contain ambiguous parts that cannot be supplemented through conversational contexts. 
CoSearchAgent will ask clarifying questions based on these ambiguous parts, thereby further refining user needs \cite{zamani2020generating, aliannejadi2020convai3}.
Specifically, guided by the task instruction $F_{clarify}$ and taking into account the multi-party conversational context $U$ along with the query $q_{rewrite}$, CoSearchAgent is tasked with asking a clarifying question aimed at gaining additional insights into the user's requirements:
\begin{equation}
    q_{clarify} = F_{clarify}(U, q_{rewrite})
\end{equation}
Here, $q_{clarify}$ represents the generated clarifying question.
If the query contains ambiguous parts, CoSearchAgent will return $q_{clarify}$ to the user and wait for the user's response. 
If not, it indicates that the query is complete, allowing CoSearchAgent to respond directly.

\subsection{Search Results Presentation}
Similar to previous collaborative search research \cite{avula2018searchbots, avula2019embedding, avula2022effects}, CoSearchAgent provides search results to all users.
As illustrated in Figure \ref{collabsearchagent}, CoSearchAgent retrieves search results relevant to the query through the search API and utilizes LLM to extract parts of each search page relevant to the query as references, thus replacing snippets in search results.

\label{extract}
\subsubsection{Fetch}
After acquiring the comprehensive and precise query $q_{rewrite}$, CoSearchAgent utilizes a search API to obtain $m$ search engine result pages:
\begin{equation}
    SERP \leftarrow \{(t_1, l_1, s_1), (t_2, l_2, s_2), \ldots, (t_m, l_m, s_m)\}
\end{equation}
Where $SERP$ stands for the search engine results pages to be retrieved, each comprising three elements: the title $t$, the link $l$, and the snippet $s$.

Subsequently, CoSearchAgent fetches the HTML content $h$ of the respective page from each provided link $l$.

\subsubsection{Extract}
While LLMs consistently face constraints on context length, the HTML contents of crawled web pages frequently surpass this limitation. To tackle this challenge, we present a two-step solution. Firstly, CoSearchAgent extracts the text portion $h_{text}$ from the HTML content $h$ of a retrieved page, serving as a representation of the current page. Secondly, utilizing the task instruction $F_{extract}$, we generate a concise summary of the current page's text content relevant to the rewritten query $q_{rewrite}$. This summary functions as a reference for the search:
\vspace{-2mm}
\begin{equation}
    ref = F_{extract}(q_{rewrite}, h_{text})
\end{equation}

Here, $ref$ denotes the reference extracted by LLM from the text content of the retrieved page. 

Ultimately, CoSearchAgent presents search results to users, wherein snippets are replaced by extracted references relevant to the query:
\vspace{-2mm}
\begin{equation}
    SERP \leftarrow \{(t_1, l_1, ref_1), (t_2, l_2, ref_2), \ldots, (t_m, l_m, ref_m)\}
\end{equation}

\begin{figure*}[htbp]
    \centering
    \includegraphics[width=\linewidth]{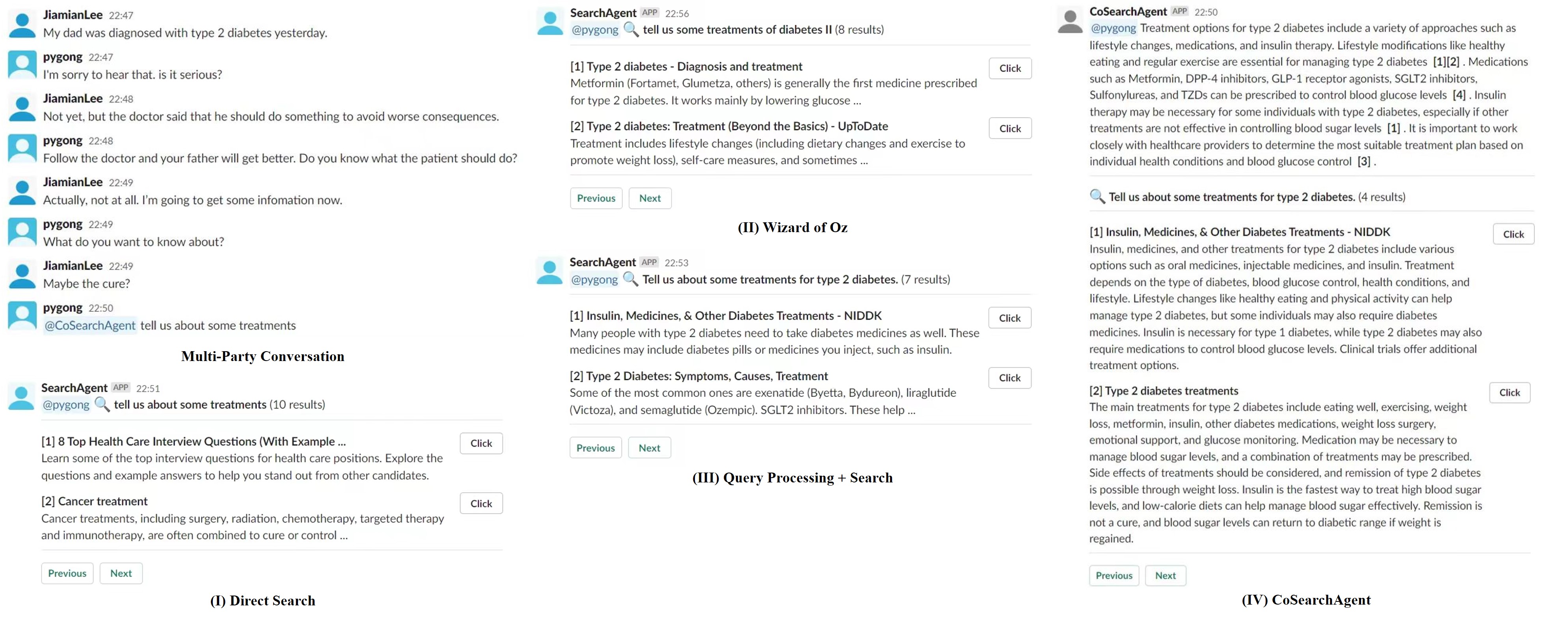}
    \caption{Example of CoSearchAgent's usage in Slack. Similar to the Wizard of Oz approach, CoSearchAgent can rewrite the query accurately for searching, and generate the accurate answer based on search results for easier user reading.}
    \label{case}
\end{figure*}

\vspace{-3mm}
\subsection{Retrieval-Augmented Generation}
Besides providing search results, CoSearchAgent also directly generates answers for users.
Some research studies indicate that utilizing references obtained through retrieval to generate answers with LLM can further enhance answer accuracy and reduce the generation of hallucinations \cite{liu2023webglm,ram2023context,vu2023freshllms}.
Therefore, we leverage references extracted from the search results as a knowledge base to generate answers.
As depicted in Figure \ref{collabsearchagent}, given the task instruction $F_{rag}$ of employing references for answering, along with the processed query $q_{rewrite}$ and a set of references $REF$ (represented as $REF = \{ref_1, ref_2, \ldots, ref_m\}$) generated by Section \ref{extract}, CoSearchAgent will produce an answer enclosed in citation marks:
\begin{gather}
    a = F_{rag}(q_{rewrite}, REF) \\
    a \leftarrow \{(seg_1, C_1), (seg_2, C_2), \ldots, (seg_k, C_k)\}
\end{gather}
Here, $a$ denotes the generated answer and can be deconstructed into $k$ components. 
Each component comprises an answer segment $seg$ and the citation marks $C$ that signify supporting references. 
Within $C$, there could be no, one, or multiple citation marks.

\section{Implementation}
We implement the CoSearchAgent plugin on Slack\footnote{https://slack.com/}, utilizing the Bolt-Python framework\footnote{https://slack.dev/bolt-python/concepts} to handle user messages and events, sending responses accordingly. 
We have designed two versions: one in English and one in Chinese.
Users can initiate queries by mentioning "@CoSearchAgent" and then CoSearchAgent will automatically respond.
In order to avoid too many interactions between CoSearchAgent and users (by clarifying questions), CoSearchAgent provides the answer and search results directly after one round of interaction. 
We utilize Serpapi\footnote{https://serpapi.com/} to access Google Search API, fetching about 10 relevant results per query and then employ request\footnote{https://github.com/psf/requests} to retrieve HTML contents, followed by html2text\footnote{https://github.com/aaronsw/html2text} for text extraction. 
Notably, search results lacking extractable references are filtered out, and CoSearchAgent generates answers using LLM itself in the absence of references.

We utilize ChatGPT\footnote{https://chat.openai.com/}, a widely used LLM configured with "temperature = 0, n = 1," to develop the CoSearchAgent, employing the "gpt-3.5-turbo-1106" version.
Due to the input length constraints of the LLM, whenever a user mentions @CoSearchAgent, it automatically captures the preceding 20 utterances as the dialogue context. 
Additionally, as the excessive length of the text content in each search result, we intercept the first 5000 tokens to enable LLM to extract query-related content.
To better harness the capabilities of the LLM, we enhance its reasoning abilities in the query processing module using the chain-of-thought method \cite{wang2022self, wei2022chain}. Additionally, in both the query processing and retrieval-augmented generation modules, we employ 5-shot demonstrations \cite{brown2020language} to reinforce the LLM's understanding of tasks and guide its output format.

Moreover, we record user behavior through three types of logs, including conversation log, search log, and click log.
as shown in Figure \ref{case}, the conversation log captures the interaction information between multiple users and CoSearchAgent.
Moreover, users can navigate up and down using the "Previous" and "Next" buttons, and search records will be logged in the Search log. Similarly, clicking the "Click" button will lead to the corresponding page, and click actions will be recorded in the Click log.
We implement log storage through MySQL\footnote{https://www.mysql.com/}.
Researchers can utilize CoSearchAgent to accomplish various collaborative search tasks, thereby obtaining user behavior logs for analysis and system optimization.

\section{Case Study}
To demonstrate that CoSearchAgent is a powerful collaborative search system, in this section, we've shown how CoSearchAgent performs on the Slack platform. 
As depicted in Figure \ref{case}, given a multi-party conversation context and a relevant user query, we offer four result-returning modes: 
(I) Direct Search: Directly search based on the user query;
(II) Wizard of Oz: Have a human read the conversational context, rewrite the query, and then search; 
(III) Query Processing + Search: LLM rewrites the query based on the conversation before searching; 
(IV) CoSearchAgent: Utilizing our plugin to return the result. 
We design a SearchAgent which accepts queries and outputs search results and the first three modes are all implemented based on it.

Results show that using the Wizard of Oz paradigm to rewrite the query allows for precise search results, whereas neglecting the conversation context can result in query failures \cite{avula2020wizard, avula2019embedding, avula2018searchbots}.
Notably, leveraging our query processing module to rewrite the query before searching also yields accurate results, and the rewritten query maintains semantic consistency obtained from the Wizard of Oz approach.
We find the best result output comes from CoSearchAgent.
CoSearchAgent not only accurately rewrites the user query for search but also offers a detailed and precise answer, supported by citation markers. 
Moreover, displaying content relevant to the query in the search results makes it more convenient for users to assess whether to navigate to the corresponding search page for detailed reading, thereby improving search efficiency.

\section{CONCLUSION}
In this demo, we introduce a lightweight collaborative search agent, CollabSearchAgent, leveraging LLMs for interactions with multiple users to fulfill their collaborative information needs on Slack.
With the capability to comprehend queries and context within a multi-user conversation and the aptitude to explore the web for pertinent information, CoSearchAgent can not only provide relevant search results but also generate accurate answers grounded on these search results to users. 
Additionally, it can seek clarification by asking questions when information needs are ambiguous.
The proposed CoSearchAgent is exceptionally flexible, making it valuable for facilitating future research on collaborative search.

\clearpage

% \begin{acks}
% \end{acks}

\bibliographystyle{ACM-Reference-Format}
\bibliography{sample-base}

\end{document}